\documentstyle[bo99,epsfig]{article}

\title{Electron injection break and \\[2mm] 
       pair content of quasar jets}

\author{M.~B{\l }a\.zejowski $^1$, M.~Sikora $^1$, R.~Moderski $^{2,1}$,
T.~Bulik $^1$ }

\affil{1) N. Copernicus Astronomical Center, Warsaw, Poland\\
2) JILA, University of Colorado, Boulder, USA}

\begin{document}

\maketitle

\begin{abstract}
We study the dependence of nonthermal radiation spectra in OVV quasars on
location of the low energy break in the electron/positron injection function.
We show that the high energy spectra produced during the outbursts
are presumably superposed from two components,
one resulting from Comptonization of emission lines, which dominates 
at MeV-GeV energies, and the other resulting from Comptonization of
infrared radiation, which dominates in the X-ray band.
\end{abstract}

\section{Assumptions}

$\bullet$ Nonthermal radiation in blazars is produced by thin shells,
propagating at a constant relativistic ($\Gamma \gg 1$) speed along the
conical jet;

$\bullet$ Relativistic electrons are injected in the shells 
within a distance range 
$\Delta r  = r_0$, starting from  $r_0$.
They are injected at a constant rate and
with the two power-law energy distribution, $Q = K \gamma^{-p}$ for 
$\gamma > \gamma_b$
and $Q \propto \gamma^{-1}$ for $\gamma \leq \gamma_b$;	

$\bullet$ Radiative energy losses of electrons are dominated by
Comptonization of the quasar broad emission lines. This process is responsible
for production of $\gamma$-rays. The low energy break at
few MeV results from inefficient radiation cooling of lower
energy electrons;

$\bullet$ Intensity of the magnetic field is $B(r)=(r_0/r) B(r_0)$.

\section{The model equations}

\subsection{Electron Evolution}

Evolution of the electron energy distribution is given by the continuity
equation (Moderski, Sikora \& Bulik 2000)
$$ {\partial N_{\gamma} \over \partial r} = - {\partial \over \partial
\gamma} \left(N_{\gamma} {d\gamma \over dr}\right) + { Q\over c\beta \Gamma} ,
\eqno(1) $$
where
$$ {d\gamma \over dr} = {1 \over \beta c \Gamma} \left(d\gamma \over
dt'\right)_{rad}- {2\over3}{\gamma \over r} . \eqno(2) $$
The second term on
the rhs of Eq.~(2) represents the adiabatic losses.
The rate of the radiative  energy losses is:

$$ \left(d\gamma \over dt'\right)_{rad} = - {4\sigma_T \over 3 m_e c } 
(u_B' + u_S' + u_{BEL}' + u_{IR}')
\gamma^2 , \eqno(3) $$
where $u_B' = B'^2/8\pi$ is the magnetic energy density, $u_S'$ is the
energy density of the synchrotron radiation field, 
$u_{BEL}' = (4/3) \Gamma^2 L_{BEL} / 4 \pi r^2 c$ is the energy density of 
the broad emission
lines, $u_{IR}' \simeq (4/3) \Gamma^2 \xi_{IR} 4 \sigma_{SB} T^4/c$ 
is the energy density of the near-IR radiation produced in molecular torus
by hot dust, and $\xi_{IR}$ is the fraction of 
the accretion disc radiation reprocessed by the dust into the near-IR band.

\subsection{Radiation Spectra}

The observed spectra as a function of time are computed using the formula
$$ \nu L_{\nu}(t) \equiv 4 \pi 
{\partial (\nu L_{\nu}(t)) \over \partial \Omega_{\vec n_{obs}}} 
= \int\!\!\!\int_{\Omega_j} 
{\nu' L_{\nu'}'[r(\theta,t), \theta]
{\cal D}^4 \over \Omega_j } {\rm d} \cos \theta {\rm d} \phi \, ,
\eqno(4) $$
where $ {\cal D} = [\Gamma(1-\beta\cos \theta)]^{-1}$ is the Doppler factor,
$\nu = {\cal D} \nu'$, and $r = c\beta (t-t_0)/(1-\beta\cos\theta)+r_0$. 
The luminosity $\nu' L_{\nu'}'$ is  contributed by: synchrotron radiation
$$ \nu' L_{S,\nu'}' \simeq 
{1\over 2} (\gamma N_{\gamma}) m_ec^2
{\left \vert{\rm d}\gamma \over {\rm d} t'\right \vert_S}  . \eqno(5) $$
the synchrotron-self-Compton (SSC) process 
%
$$ \nu'L_{SSC, \nu'}' = {\sqrt 3 \sigma_T \over 8 \Omega_j r^2} {\nu'}^{3/2}
\int N_{\gamma} \left [\gamma= \sqrt {3 \nu' \over 4
\nu_S'} \right ] L_{S,\nu'}' \nu_S'^{-3/2} \, {\rm d} \nu_S' \,
, \eqno(6) $$
and the external-radiation-Compton (ERC) process
$$\nu' L_{ERC,\nu'}' [\theta'] \equiv 
4 \pi {\partial  (\nu' L_{ERC,\nu'}') \over \partial \Omega_{\vec n'_{obs}}'}
\simeq {1\over 2} \gamma N_{\gamma} 
m_e c^2 
\, \left \vert {\rm d} \gamma \over {\rm d} t' \right \vert_{ERC} [\theta']
\, , \eqno(7) $$
where
$$ \left\vert {\rm d} \gamma \over {\rm d} t'\right \vert_{ERC} [\theta'] 
\simeq 
{4 \sigma_T \over 3 m_e c^2} \gamma^2 {\cal D}^2 (u_{BEL} + u_{IR})
 \, . \eqno(8) $$
and 
$$ \nu' \simeq {\cal D} \gamma^2 \nu_{BEL/IR} \, , \eqno(9) $$
where $\nu_{BEL}$ and $\nu_{IR}$ are average frequencies of broad
emission lines and of infrared radiation of hot dust, respectively.
Note that in the comoving  frame the ERC radiation field is
anisotropic, while  the synchrotron and SSC radiation fields
are isotropic (Dermer 1995).
%
 
\begin{figure}[h]
\vspace{-0.2cm}
\centerline{\psfig{file=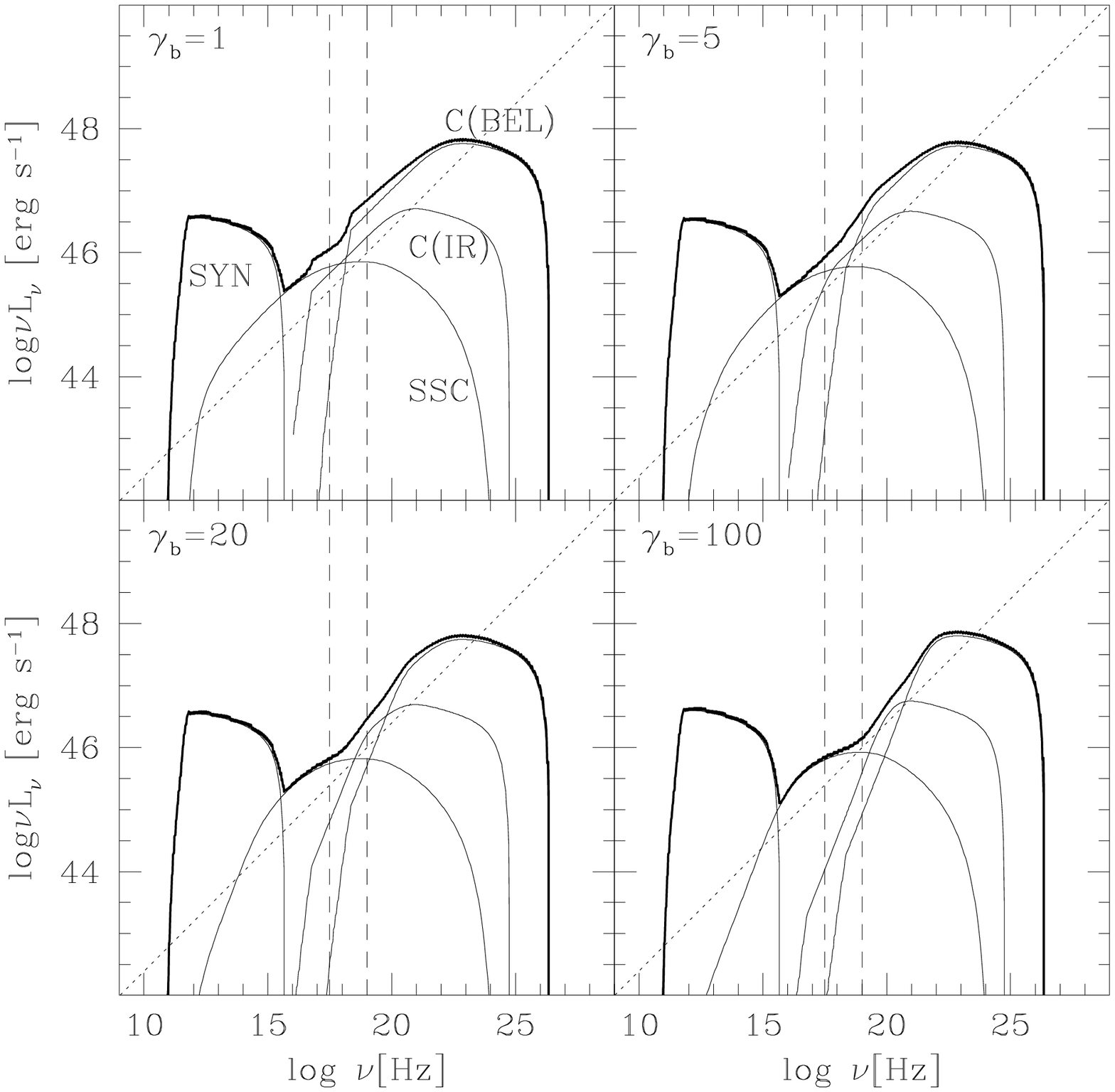,width=0.8\textwidth}}
\vspace{-0.6cm}
\caption[]{The time averaged blazar spectra  for four values of
$\gamma_b$. The dotted line marks the typical slope of the X-ray
spectra in blazars  ($\alpha=0.6$);
the dashed lines enclose the 1-30 keV X-ray band.
In each panel we show four spectral components: synchrotron
(SYN), synchrotron-self-compton (SSC), comptonization of broad
emission lines [C(BEL)], and comptonization of near-IR dust
radiation [C(IR)].
}
\end{figure}

\section{Results}

 In Figure 1 we present the  time averaged blazar spectra for
four different values of the energy break $\gamma_b$. All models
are computed for the following set of parameters:  $r_0 = 6
\times 10^{17}$ cm; $\Gamma=15$; $L_{BEL}=1.4 \times 10^{44}$
ergs/s; $B'= 1.4$ Gauss; $\gamma_{max}=10^4$; $p=2.2$; $K= 0.7
\times 10^{50}$ s$^{-1}$;  $\theta_{obs} = \theta_j=1/15$\,rad,
$T=1000$ K, and $\xi_{IR}=0.08$. For the justification of this
choice see B{\l}a{\.z}ejowski et~al (in preparation).

\section{Discussion}

\subsection{Low Energy Break in Electron Distribution}

We can see from Fig. 1, that for $\gamma_{b}=1$ the low energy tail of 
the C(BEL)
component extends down to $\sim 2$ keV. Thus, any presence of thermal
nonrelativistic electrons should be imprinted as a bump, peaking around 2 keV. 
Since blazar spectra extend down to much lower values without any
bump  (Comastri et al. 1997; Sambruna 1997; Lawson \& McHardy 1998), 
we exclude the domination of C(BEL) in the soft X-ray band.  

In order to get  soft X-ray spectra which smoothly join the middle X-ray band,
one needs to assume that $\gamma_{b} >3$. Then 
the low energy break of C(BEL) moves above  $>20$ keV and X-ray radiation below
this value is dominated by either by SSC or C(IR). Noting 
that the  SSC X-ray spectra are much 
softer than the observed ones ($\alpha_{X,SSC} \sim 1$ vs. 
$\alpha_{X,obs} \sim 0.6-0.7$; Kubo et al. 1998), C(IR) is a 
better candidate for X-ray production. 
This, however, can be the case if $\gamma_{b} \le 10$. For larger 
values of $\gamma_{b}$ the low energy break of the C(IR) component  moves
to energies $> 20$ keV, and, then, at lower energies the C(IR) spectrum   
becomes too hard in comparison  with observations.
We conclude that interpreting the blazar X-ray observations within the
framework of our model implies that $\gamma_b$ is enclosed in the range 
$(3 - 10)$. 


\subsection{Pair Content}

For a jet dynamically dominated by the energy flux of protons, $L_p$, and for
radiative energy losses of electrons dominated by Comptonization of
broad emission lines, the pair content
of the jet can be calculated from the formula (Sikora et al., in preparation)
$$ {n_{pairs}' \over n_p'} \sim {K \over 2(p-1)\gamma_{b}^{p-1}} 
{m_p c^2 \over L_p} \Gamma^2  \, ,
\eqno(10) $$
This, for our model parameters and $3 < \gamma_{b} < 10$ gives
$$ 6/L_{p,47} < n_{pairs}' / n_p' <  26/L_{p,47} \, .\eqno(11) $$
Thus, our results suggest that particle number in quasar jets is
dominated by pairs, while the jet inertia is still dominated by
protons.

\begin{acknowledgements}
This project was supported by ITP/NSF grant PHY94-07194, the Polish KBN 
grant 2P03D00415, and NASA grant NAG-5-6337.
\end{acknowledgements}

\end{document}